# Is Bonferroni correction more sensitive than Random Field

# Theory for most fMRI studies?

Tim M. Tierney[a], Christopher A. Clark[a] , David W. Carmichael[a]

[a] UCL Institute of Child Health, University College London, London, UK.

Corresponding Author: Tim M. Tierney
Address: 30 Guilford Street, UCL Institute of Child Health,
University College London, London, UK.
Phone Number: +44 (0)20 7905 2130
Email: tim.tierney.12@ucl.ac.uk

**Abstract**

Random Field Theory has been used extensively in the fMRI literature to address the multiple comparisons problem. The method provides an analytical solution for the computation of precise p-values when its assumptions are met. When its assumptions are not met the thresholds generated by Random Field Theory can be more conservative than Bonferroni corrections, which are arguably too stringent for use in fMRI. As this has been well documented theoretically it is surprising that a majority of current studies (~80%) would not meet the assumptions of Random Field Theory and therefore would have reduced sensitivity. Specifically most data is not smooth enough to meet the good lattice assumption. Current studies smooth data on average by twice the voxel size which is rarely sufficient to meet the good lattice assumption. The amount of smoothing required for Random Field Theory to produce accurate p-values increases with image resolution and decreases with degrees of freedom. There is no rule of thumb that is valid for all study designs but for typical data (3mm resolution, and greater than 20 subjects) residual smoothness with FWHM = 4 times voxel size should produce valid results. However, it should be stressed that for higher spatial resolution and lower degrees of freedom the critical smoothness required will increase sharply. This implies that researchers should carefully choose appropriate smoothing kernels. This can be facilitated by the simulations we provide that identify the critical smoothness at which the application of RFT becomes appropriate. For some applications such as presurgical mapping or, imaging of small structures, probing the laminar/columnar structure of the cortex these smoothness requirements may be too great to preserve spatial structure. As such, this study suggests developments are needed in Random Field Theory to fully exploit the resolution of modern neuroimaging.

# 1. Introduction

The question of how to address the issue of multiple comparisons in fMRI has received a lot of attention during recent years. Bennet et al (2009) suggested that a large number of studies were using thresholds that were uncorrected for multiple comparisons. Furthermore, Bennet et al (2011) highlighted the inadequacy of uncorrected thresholds when they demonstrated that post-mortem images of a salmon could display "activation" when multiple comparisons were not controlled for. They concluded that there was a need in the imaging literature for more adequate control of statistical error rates.

The key difference between the multiple comparisons issue in traditional statistics and imaging is the presence of spatial correlation. Spatial correlation makes methods such as Bonferroni correction inappropriate for the control of statistical error rates in imaging as they are too conservative. Topological inference was introduced to address this issue using the theory of stochastic processes (Friston et al 1992) and the Euler characteristic (Worsley et al 1993), which is often referred to as Random Field Theory (RFT) Worsley (1996). The advantage of RFT is that it recognises that data is sampled from a continuous field. This means that researchers can make inference on topological features (and not voxels) at arbitrary resolution.

This method is computationally inexpensive and accurate when its assumptions are met. Unfortunately the mathematical complexity of the method means that it is difficult to precisely test the assumptions necessary for its application However, a thorough and accessible review of the assumptions can be found in Petersson et al (1999). This lack of clarity surrounding the assumptions concerning RFT is evident from the fact that few studies, in the fMRI literature have explicitly tested and reported whether or not they have met all the assumptions of RFT before utilising the method. Therefore in each case the validity of RFT is implicitly assumed.

This can potentially explain the large number of studies that have reported that RFT can be conservative (Roels et al, 2015; Li et al, 2015; Li et al, 2014; Durnez et al, 2014; Pantazis et al, 2005; Worsley, 2005; Hayasaka et al, 2004; Nichols & Hayasaka, 2003; Hayaska & Nichols, 2003; Worsley, 2003, Eklund et al, 2016). The conservative nature of RFT in these scenarios is most likely due to situations where the good lattice assumption is not met as these studies largely report a smoothness dependence on the results. The good lattice assumption has been stated by Flandin and Friston (2015) as the following:

"The component (error) fields conform to a reasonable lattice approximation of an underlying random field with a multivariate Gaussian distribution."

This statement can be broken down into two sub-statements. The first is that the component (error) fields have a multivariate Gaussian distribution. This is a testable assumption using Mardia's test for multivariate normality (Barnes et al, 2013). As such, we will not consider further how departures from multivariate normality may hinder inferences using RFT.

However, the second statement that the components (error) fields conform to a reasonable lattice approximation to an underlying random field is not as easily testable. This statement can be intuitively understood as meaning the data needs to be sampled sufficiently to represent the topological features of interest. This can be facilitated by smoothing the data so the topological features (the Blobs) become large relative to the voxel size and therefore well sampled. The question remains as to how smooth the data needs to be to meet this assumption.

The literature is quite variable concerning advice on this matter. Smoothing is recommended to be between 3-5 times the voxel size for RFT to be accurate (Petersson et al, 1999; Worsley, 2003; Barnes et al, 2013; Worsley, 2005). This ambiguity in the literature means that it is difficult for a researcher to assess if their dataset has met this assumption. This is quite problematic because if this assumption is not met then the thresholds produced by RFT will be more conservative than a Bonferroni correction (Worsley, 2005). Considering this ambiguity in the literature concerning this assumption and lack of suitable tests for its validity this study poses the following question.

1. Do current data analysis and acquisition strategies produce thresholds more stringent than Bonferroni?

To answer this question, we performed simulations using parameters derived from a survey of recent fMRI studies.

## 2. Theory

It is possible to argue that RFT and Bonferroni should not be compared as RFT is designed to work in continuous space and Bonferroni in discrete space. This fundamental difference means that RFT theory derived thresholds do not change by image resampling whereas a bonferroni will as it is based on the number of voxels

However, one can use the Bonferroni correction to establish the lower bound on smoothness above which RFT provides accurate p-values. In other words, by decreasing the spatial resolution of images, (or increasing voxel size for a fixed smoothness) there will be a point at which the RFT correction becomes more conservative than the Bonferroni correction. For the purposes of this study we use this point as our definition of when the good lattice assumption is violated. This definition is chosen for practical reasons as it defines regimes of smoothness (or voxel size) in which it is and is not appropriate to apply RFT and is therefore of practical relevance to the imaging community.

## 3. Methods

### 3.1 Analysis

### 3.1.1 fMRI Survey Descriptive Statistics

Brain volume data is taken from a meta-analysis of brain volume (Borzage et al 2014). Voxel size and smoothness are taken from fMRI studies published in *NeuroImage and NeuroImage:Clinical* between January 1[st] and February 25[th]. This included articles in press. For the survey we assume that papers published in NeuroImage are a representative sample of the data analysis and acquisition practices of imaging researchers and therefore allow us to make an appropriate generalisation of current practice.

Using Science direct 198 studies included the word fMRI. Only 137 were included in the analysis due to the following reasons: 1) not all studies reported smoothness and voxel size, 2) they were simulations, 3) they were introducing software/repositories, 4) they were animal studies, 5) they weren't fMRI studies but mentioned fMRI (fNIRS, optogenetics), 6) They were reviews/meta analyses.

The 137 studies used a variety of different error control methods. These were corrected Parametric (68/137 = 49.6%), uncorrected Parametric (24/137 = 17.5%), Simulation based corrections (17/137 =12.4%, machine learning (8/137=5.8%) FDR (6/137 = 4.4%), Threshold Free Cluster enhancement (3/137 =2.2%), Non-parametric permutation (1/137 = 0.73%), mixture modelling (1/137 = 0.73%), Bonferroni (1/137 = 0.73%), not reported (8/137 = 5.8%). One might expect that as only the corrected Parametric approaches (RFT) make assumptions concerning the image smoothness the smoothness may be different between studies that use RFT and those that do not. This was not found to be the case (using welch's two sample t-test). Studies that use RFT on average smooth their data by 2.05 times voxels size where as those that does not use RFT smooth by 1.94 times voxel size( $t$ (133.85) = 1.0225, $p$ = .3084, effect size: $r$ = .088 ). As such, using values of smoothing obtained from all of these studies is justified when trying to infer the appropriateness of RFT for current image acquisition and analysis strategies.

As no study reported the estimated residual smoothness from their analysis, which is crucial to determining the appropriateness of RFT, we only documented the applied smoothing kernel width. This is an underestimate of the smoothness of the component error fields as images are already slightly smooth due to the point spread function of the image, T2* blurring and post processing techniques such as interpolation which also increase smoothness. We account for this in our analysis by using empirical data to describe the relationship between estimated residual smoothness and applied smoothing kernel width.

This empirical data is drawn from 53 EEG-fMRI studies of focal epilepsy patients (for the purpose of localising the epileptic focus) conducted in the data's native space. (Centeno et al, 2016). The fact that these studies were conducted in native space is important because the process of normalisation/nonlinear warping alters smoothness in an algorithm specific fashion which we do not intend to investigate here. Having calculated the ratio of estimated residual smoothness to applied smoothing kernel width we use the 5[th], 50[th] and 95[th] quantiles to quantify the uncertainty in this relationship.

### 3.1.2 Simulation of RFT thresholds and Bonferroni Thresholds

In order to simulate RFT thresholds smoothness was defined relative to voxel size (e.g. if voxel size is 3mm and FWHM = 9 the smoothness was 3) and was varied between 1 and 6. The degrees of freedom were varied between 10 and 100. This was repeated for 1mm, 2mm and 3mm isotropic voxels. This simulation was performed using t-fields designed to achieve FWE correction at p<0.05. We make the following simplification for calculating the *Resel Count.*

$$Resel\ Count = \frac{Volume}{FWHM_x FWHM_y FWHM_z} \qquad \textbf{1}$$

While this is not strictly true it is a reasonable approximation for large unmasked volumes (Worsley et al, 1996). Furthermore, if all the components that constitute the *Resel Count* were to be included in the analysis RFT thresholds would be even more conservative (see Worsley (1996) for a description of these other parameters). It is also worth noting that the Volume can be in mm$^3$ or cubic voxels as long as the FWHM is measured in the same unit. The Bonferroni correction was calculated based on the number of voxels, for a given resolution, that would fit in a whole brain volume of 1.4 litres.

### 3.1.3 Comparison of Theory and Practice

Having established when RFT would produce overly conservative thresholds (from the simulation described in Section 3.1.2) and given the typical values of smoothness and voxel sizes found in the literature  (found in the survey described in Section 3.1.1) we can now assess how likely a given study is to meet the assumptions of RFT.  By taking the ratio of the smoothing kernel width to the voxel size (both found from the survey) we create a new distribution. This new variable (the ratio of smoothness to voxel size) can be compared to the simulation described in section 3.1.2 to see how many studies reach the critical smoothness required for the successful application of RFT.

To account for the limitation of unknown residual smoothness we assume a multiplicative relationship between residual smoothness and smoothing kernel width. We use the empirical quantiles described in section 3.1.1 to get more realistic estimates of residual smoothness. As we will show that the distribution of applied smoothing kernel width relative to voxel sizes is normal the adjusted distribution, obtained using the empirical quantiles, has an analytical form that is obtained by simply multiplying the mean and the standard deviation of the original distribution by the quantiles described in section 3.1.1 to calculate a an upper bound, measure of central tendency and lower bound on the number of studies likely to meet the god lattice assumption.

## 3.2 Software

All probabilities are computed using the R programming language (R Core team, 2015). The normal distribution of the ratio of smoothness to voxel size was fit using the fitdistrplus package (Delignette-Muller & Dutang, 2015). The RFT thresholds are computed using SPM12 ([www.fil.ion.ucl.ac.uk](www.fil.ion.ucl.ac.uk)).

## 4. Results

## 4.1 Descriptive Statistics

Of the 137 studies reviewed the average voxel size in the x and y direction is 3.01 mm (SD = 0.62 mm). In the slice direction the average slice thickness was found to have a mean of 3.53 mm (SD= .080 mm). The average FWHM of smoothing kernels used was 6.12 mm (SD =2.11 mm). The histograms describing these variables are presented in Figure 2

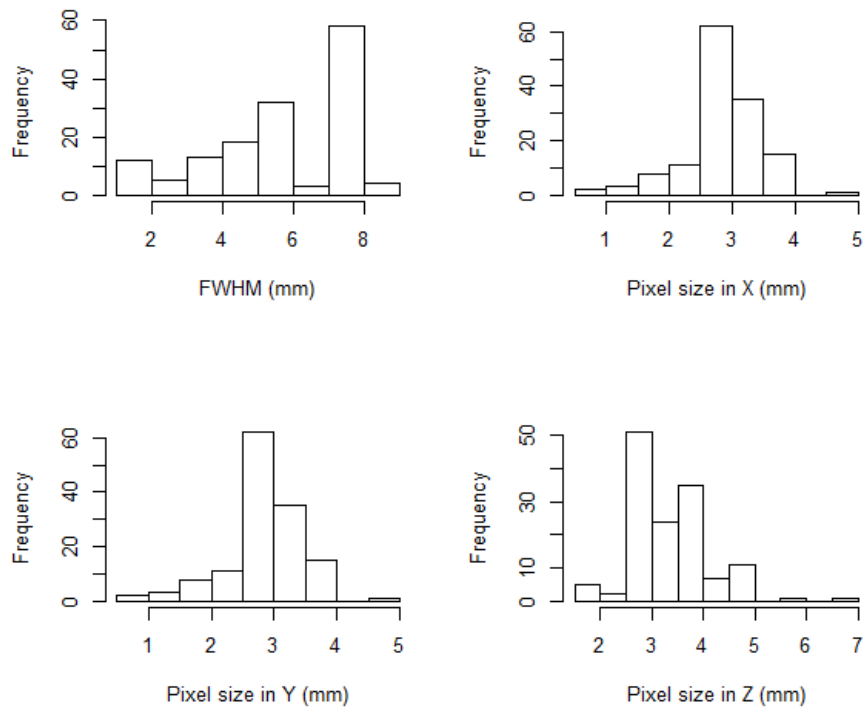

*Figure 2.* Histograms of voxel sizes and smoothing kernel widths.

From Figure 2 it is clear that the observed distributions are complex and do not obviously conform to easily describable probability distributions. In particular the histogram of smoothing kernel width is interesting as it shows a clear peak at 8mm which corresponds to the SPM (89/137 studies used SPM) default smoothing kernel width.

The ratio of estimated smoothness to applied smoothing kernel width from Centeno et al (2016) had a .05 quantile = 1.26, median (.5 quantile) = 1.36 and .95 quantile = 1.77. The distribution is graphically represented in Figure 2. We can use these descriptive statistics to create approximate upper and lower bounds on the analysis presented in section 4.3.

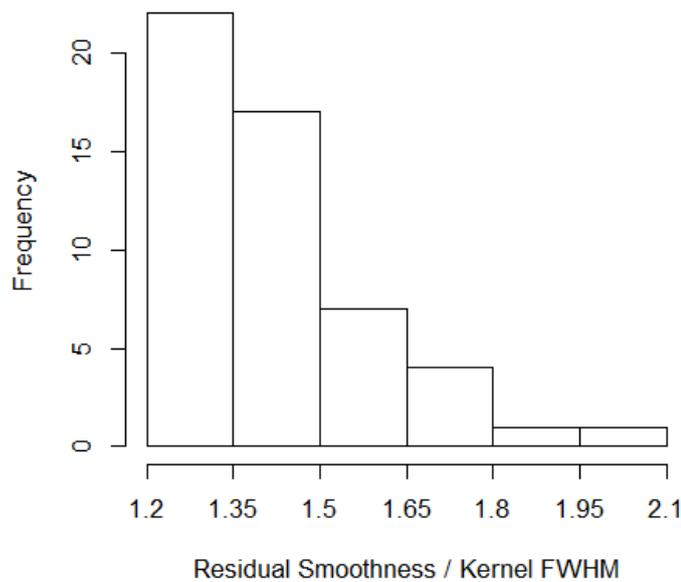

*Figure 3.* Histogram of estimated residual smoothness (i.e. the smoothness of residual fields – as opposed to signal) relative to applied smoothing kernel width in a sample of N=53 subjects taken from Centeno et al (2016).

## 4.2 When does RFT produce less conservative thresholds than Bonferroni?

To address this question we compare RFT theory thresholds to the Bonferroni threshold (black lines in Figure 4) for different degrees of smoothness and degrees of freedom using a t-field. The results are represented graphically in Figure 4. The black lines in Figure 4 indicate the location of the critical smoothness threshold at which RFT fails. As expected there is strong dependence of the critical threshold on the degrees of freedom but there is a more surprising dependence on resolution as well. High resolution data needs to be smoothed relatively more in order to meet the good lattice assumption: the black lines are shifted to the right indicating larger relative smoothness is required. These results show that there is no rule of thumb that is valid for all experimental designs. The smoothness requirements to meet the good lattice assumption are study specific.

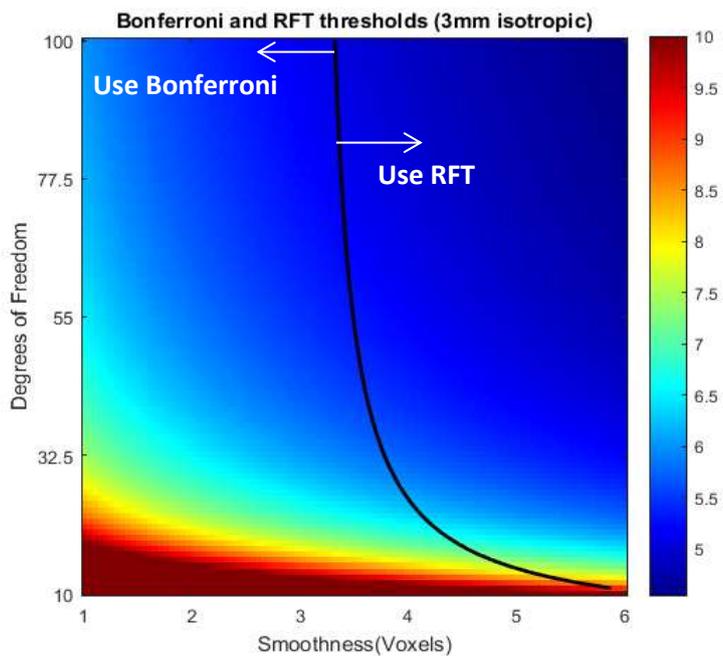

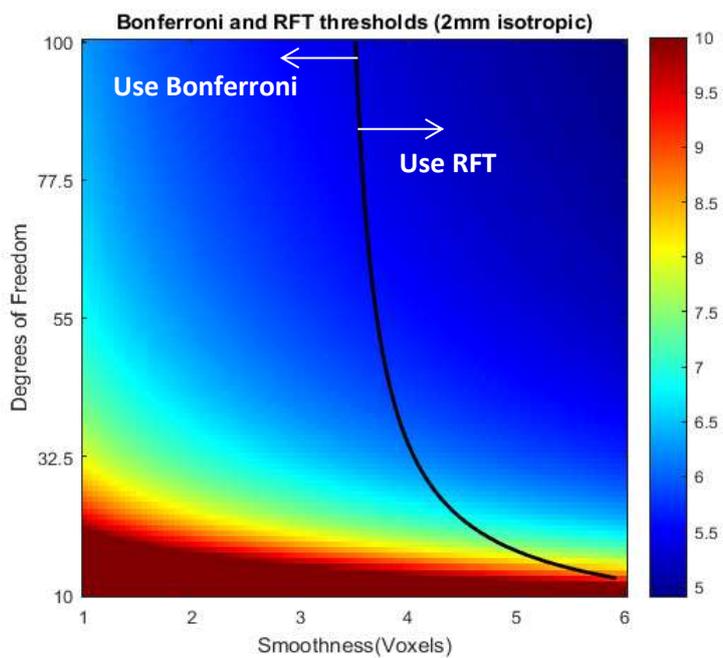

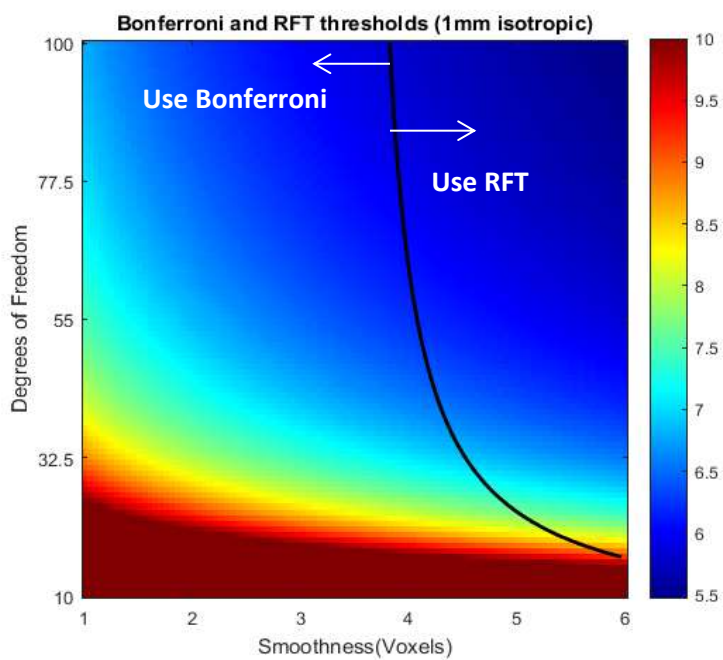

*Figure 4.* Sensitivity of RFT to degrees of freedom and smoothness. The x axis represents the smoothness relative to the voxel size. The y axis displays the degrees of freedom. The colour bar encodes the height of the t-statistic required to achieve FWE p<.05 using RFT on a brain that is 1.4 litres in volume. The black line shows where the RFT threshold is equal to the Bonferroni threshold (where RFT starts producing accurate thresholds). To the right of the black line RFT is less conservative than Bonferroni and to the left of the black line RFT is more conservative than bonferroni. This simulation is repeated for 3mm, 2mm and 1mm voxels. For display purposes colour bar have been capped at t=10.

**4.3 Do current data analysis and acquisition strategies meet the assumptions of RFT?**

The histogram of the applied smoothing kernel width relative to voxel size is displayed in Figure 5. Empirical and theoretical quantiles, probabilities and a cumulative density function accompany this histogram to illustrate the reasonably close fit of this variable to a normal distribution (The closer the points are to the lines the more appropriate the assumption of normality is). The mean of this distribution = 1.99 voxels (SD =.64 voxels).

Using this distribution we can predict the probability of a study having smoothness greater than 3.5 (the point in Figure 4 where the degrees of freedom dependence begins to vanish in the lowest resolution condition) times the voxel size – therefore satisfying the RFT assumptions. The probability is .009 corresponding to a less than 1% chance of a study fulfilling RFT assumptions and obtaining a threshold less conservative than bonferroni.

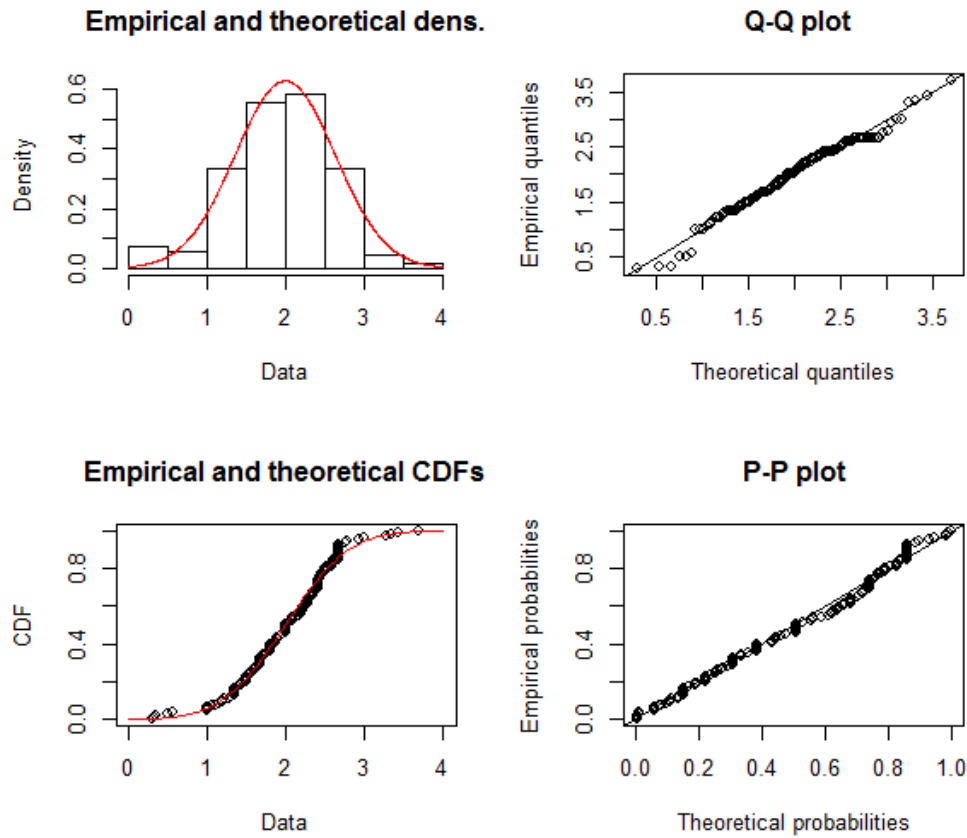

*Figure 5.* Normality of FWHM / voxel size. The empirical values for FWHM/voxel size are compared with theoretical values for a normal distribution of mean = 1.99 voxels, SD =.64 voxels. As the empirical, quantiles, probabilities and cumulative density function are a good fit to their theoretically predicted values the assumption of normality of FWHM/voxel size is reasonable.

This is however an overestimate of the numbers of studies that fail to meet the assumptions of RFT – as it is based on smoothing kernel width and not the residual smoothness. However, using the empirical bounds described in Section 3.1.1 we can adjust the distribution in Figure 5 and recompute the probabilities. This produces a more realistic probability of a study being sufficiently smooth enough to meet the assumptions of RFT.

Adjusting by the multiplicative factors in Section 4.1 the median prediction of the percentage of studies likely not to meet the assumptions of RFT is 82%. The upper and lower bounds for the prediction derived using the .95 and .05 quantile described in section 3.1.3 and computed in section 4.1 are 49% and 89% (obtained by multiplying mean and standard deviation by 0.95 and 0.05 quantiles respectively). Therefore the majority of studies are unlikely to meet the assumptions of RFT. This is graphically represented in Figure 6

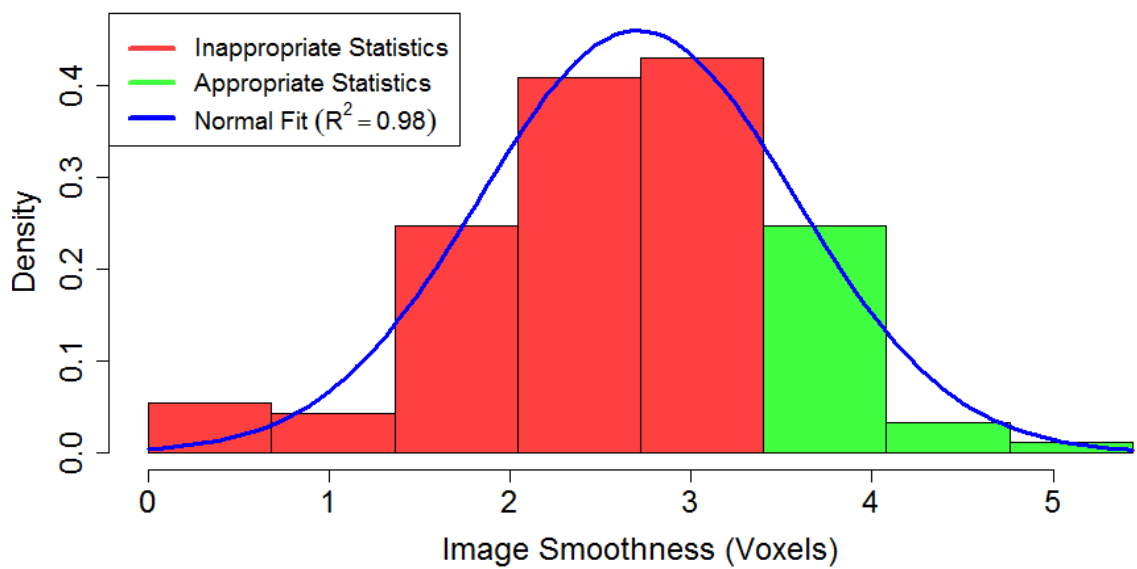

Figure 6. Estimate of residual smoothness in current studies. In this figure the distribution of smoothness to voxel size is re-estimated accounting for relationship between residual smoothness and applied smoothing kernel width. The figure demonstrates that ~80% of studies are unlikely to meet the good lattice assumption in the most lenient case of low resolution data (3mm isotropic) where the criteria for meeting the good lattice assumption is that smoothness be ~3.5 times voxel size.

## 5. Discussion

### 5.1 Summary

We have demonstrated that there is no rule of thumb regarding how smooth ones data should be that can account for the variability in study designs found in the literature. However researchers can use Figure 4 to identify the critical smoothness necessary to apply RFT in a meaningful way. Although it should be noted this is a lower bound on the smoothness that should be applied to the data and the threshold will still be conservative just less so than a bonferroni correction. In reality, to achieve optimal control of FWE the data may need to be smoothed even further (Worsley, 2005) .These figures are important to interpret the empirical data we have reviewed. We have presented empirical data that shows that – on average – researchers smooth by twice their voxel size and very few are likely to reach the minimal smoothness required  to meet the assumptions of RFT. In short for the majority (~80%) of current fMRI studies bonferroni corrections are more sensitive than RFT.

## 5.2 Alternate Solutions

If researchers do not wish to smooth their data sufficiently to meet the assumptions of RFT (which was found to be the case in this study) a number of other options exist. Worsley (2005) provided an initial framework for addressing these issues but it required numerical integration, was inaccurate for t and F fields with low degrees of freedom and requires that the autocorrelation function of the field be Gaussian. Furthermore, the results were often conservative for some values of smoothness. The restriction on the shape of the autocorrelation function means that structurally adaptive smoothing kernels would violate this method's assumptions (Andrade et al, 2001). These kernels are crucial to improving spatial specificity in the high resolution imaging that is increasingly available (Yacoub et al, 2008).

Different error rates could be controlled such as the False Discovery Rate (FDR) as described by Genovese et al (2002). While this is a reasonable form of error control the continuity inherent in brain imaging data means there are a number of situations where the direct application of FDR (although it does not make any assumptions on image smoothness) is limited. Work by Chumbley and Friston (2009) has sought to address these issues but relies on the use of theoretical results from RFT, which require smooth data

Alternate approaches to control for FWE include non-parametric permutation tests. These methods make minimal assumptions about the data and are suitable for group-level studies regardless of smoothness, voxel size and degrees of freedom (unlike RFT). The cost incurred is one of computational burden and difficulties in construction of models with covariates and nuisance variables (Winkler et al, 2014). Furthermore the application of permutation tests to the analysis of individual subjects (important in clinical applications) is difficult as the assumption of exchangeability is often violated due to autocorrelation in the data. This is not a trivial issue as modern fMRI sequences can reach repetition times of 100ms in clinical applications (Jacobs et al, 2014). Some work has been done to address this issue but requires knowledge of the autocorrelation function (Adolf et al, 2014).

In practice some software such as SPM simply choose the minimum of the bonferroni correction and the RFT correction. While this is a perfectly reasonable approach to prevent the use of needlessly strict thresholds it should be noted that the bonferroni correction itself is quite a stringent threshold and inappropriate for smooth data (Worsley, 2005).

## 5.3 Implications

In theory the use of RFT for topological inference seems a reasonable choice for controlling FWE in any spatial dataset as it allows for inference that is independent of the native resolution, has an analytical solution for many different statistical fields, which can be computed quickly, that is highly accurate for smooth fields of arbitrary dimensionality and geometry.

However, in spite of these benefits our simulations and empirical data presented seem to suggest that very few current studies would benefit from the use of RFT; indeed the thresholds generated are more conservative than a Bonferroni correction. This is because the RFT derived p-values are valid only for continuous fields. Measured data is always discrete but can approximate continuous data if smoothed sufficiently. In short, the reviewed data is not smoothed sufficiently.

As the majority of studies in the fMRI literature are group-level studies the situation becomes slightly more complex as one cannot assume the degrees of freedom are large. In fact when degrees of freedom are less than 30 (see Figure 4) the smoothness requirements begin to increase rapidly. It is therefore difficult to know how much inference is hampered at the group level due to varying sample sizes and nonlinear warping methods (which will influence smoothness in an algorithm specific fashion). As such it is crucial that residual smoothness values are reported and smoothing kernels are chosen carefully with respect to these factors if RFT is used. In these situations, if the assumptions are not met non-parametric inference may be a reasonable alternative so as to ensure validity (Winkler et al, 2014).

There are a number of situations where it is not feasible to perform group analyses. In these cases there are very few alternatives to RFT for the control of FWE rates. These include clinical applications such as the presurgical mapping of epilepsy patients (Duncan et al, 2016) or basic science applications involving small structures. In particular researchers are now interested in probing the layered/columnar structure of the cortex using high resolution fMRI which must be done in the single subject space (Yacoub, 2009; Heinzle et al, 2016).

The increased capability to image at high resolution brought about by the availability of MRI scanners with increased magnetic field strength and high density receive array coils means that there is a much greater need for methods that can control for multiple comparisons in this context. It is therefore necessary for developments in RFT to explicitly incorporate the sampling of the continuous field (the brain) in situations where image smoothness needs to be kept at a minimum – and may be nonstationary and anisotropic (as is the case with Voxel Based Morphometry).

RFT would then be bounded by Bonferroni (so that RFT thresholds are never higher than Bonferroni thresholds) and make them appropriate with less stringent requirements for smoothing (which is crucial for imaging at higher resolution). This would allow for the application in single subject analysis and high resolution imaging. Some work has been done to address this issue using simulations (Li et al, 2015; Li et al, 2014) but it is not an analytical solution and as such it is difficult to treat theoretically or extend the results to other statistical fields.

## 5.4 Limitations

In section 4.3 it is suggested that 80% of current studies do not meet the assumptions of RFT at an individual level. However, we had to approximate residual smoothness as it is seldom reported and therefore could not be incorporated into our analysis. We finessed this limitation by assuming a multiplicative relationship between estimated smoothness and applied smoothing kernel and created bounds on this relationship with empirical data at the individual level. While this is not optimal, it is likely to be a reasonable approximation.

For simplicity the simulation in section 4.2 does not account for dependence on brain volume or geometry. The implications for cluster based inference are also not considered here but are considered elsewhere (Eklund et al, 2016; Flandin & Friston, 2016). This is because the purpose of this study was not to provide a complete validation of RFT but to highlight this issue of reduced sensitivity researchers may face when trying to interpret their data at an individual level.

## 6. Conclusions

We have argued that the Bonferroni correction provides a crucial point of reference that identifies a critical bound on smoothness (or voxels size), which permits the use – or not – of RFT. We have further shown that it is impossible to generate a rule of thumb that could guide researchers on how much smoothing should be applied to their data considering the variability in study designs. They must instead carefully choose the kernel to account for their voxel size, degrees of freedom (at the single subject level and group level) and registration routines (although for the "average" study with 3mm resolution and more than 20 subjects a FWHM = 4 times voxel size may suffice). While the effects of smoothing have been previously documented we present evidence suggesting most published studies this year (80%) do not meet the assumptions of RFT. However this inference is limited by the lack of reported estimated smoothness values in the literature. This information is crucial to understanding the validity of statistical thresholds and should always be reported. Future work is required in RFT to explicitly incorporate the sampling of continuous fields in order to fully exploit the ever increasing spatial precision of fMRI data.

## Acknowledgements

The author's would like to thank Karl Friston, Guillaume Flandin and the FIL method's group for helpful advice in the conceptualisation of this paper. Research presented in this paper was funded by University College London IMPACT, the James Lewis Foundation via the Great Ormond Street Hospital Children's Charity, the Child Health Research Appeal Trust and Action Medical Research grant number SP4646. This work was undertaken at Great Ormond Street Hospital/UCL Institute of Child Health who received a proportion of funding from the UK Department of Health's NIHR Biomedical Research Centres funding scheme.